\documentclass[12pt,fleqn]{article}
\usepackage[DIV12]{typearea}
\usepackage{amsmath,amsfonts,amssymb}
\usepackage{graphicx}
\usepackage{cite}
\usepackage{mathrsfs}
\usepackage{bbm}
\usepackage{pifont}
\usepackage{cancel}
\usepackage{units}
\usepackage[small,loose]{subfigure}  
\usepackage{color}
\usepackage{psfrag}

\newcommand{\Secref}[1]{Section~\ref{#1}}
\newcommand{\Eqref}[1]{Equation~\eqref{#1}}

\newcommand{\Tabref}[1]{Table~\ref{#1}}

\newcommand{\eVdist}{\kern-0.06em}

\newcommand{\rep}[1]{\ensuremath\boldsymbol{#1}}
\newcommand{\crep}[1]{\ensuremath\overline{\boldsymbol{#1}}}

\DeclareMathOperator{\adj}{adj}

\newcommand{\D}{\mathrm{d}}
\newcommand{\I}{\mathrm{i}}

\newcommand{\E}[1]{\ensuremath{\mathrm{E}_{#1}}} 

\newcommand{\SO}[1]{\ensuremath{\mathrm{SO}(#1)}}
\newcommand{\SU}[1]{\ensuremath{\mathrm{SU}(#1)}}
\newcommand{\U}[1]{\ensuremath{\mathrm{U}(#1)}}
\newcommand{\Z}[1]{\ensuremath{\mathbbm{Z}_{#1}}} 

\newcommand{\hu}{\ensuremath{H_u}}
\newcommand{\hd}{\ensuremath{H_d}}
\newcommand{\qhu}{\ensuremath{q_{\hu}}}
\newcommand{\qhd}{\ensuremath{q_{\hd}}}

\newcommand{\lle}{\ensuremath{\left(L\,L\,\overline{E}\right)}}
\newcommand{\lhde}{\ensuremath{\left(L\,\hd\,\overline{E}\right)}}
\newcommand{\lhu}{\ensuremath{\left(L\,\hu\right)}}

\allowdisplaybreaks[1]

\numberwithin{equation}{section}

\def\mytitle{The $\boldsymbol{\mu}$ term and neutrino masses}

\title{\mytitle}

\begin{document}

\begin{titlepage}

\begin{flushright}
UCI--TR--2012--08\\
TUM--HEP 845/12\\
DESY-12-108\\
FLAVOUR(267104)-ERC-19
\end{flushright}

\vspace*{1.0cm}

\begin{center}
{\Large\bf
\mytitle
}

\vspace{1cm}

\textbf{
Mu--Chun Chen\footnote[1]{Email: \texttt{muchunc@uci.edu}}{}$^a$,
Michael Ratz\footnote[2]{Email: \texttt{michael.ratz@tum.de}}{}$^b$,
Christian Staudt\footnote[3]{Email: \texttt{christian.staudt@tum.de}}{}$^b$,
Patrick K.S.~Vaudrevange\footnote[4]{Email: \texttt{patrick.vaudrevange@desy.de}}{}$^c$
}
\\[5mm]
\textit{\small
{}$^a$ 
Department of Physics and Astronomy, University of California,\\
~~Irvine, California 92697--4575, USA
}
\\[5mm]
\textit{\small
{}$^b$ Physik--Department T30, Technische Universit\"at M\"unchen, \\
~~James--Franck--Stra\ss e, 85748 Garching, Germany
}
\\[5mm]
\textit{$^c$\small~Deutsches Elektronen--Synchrotron DESY, Notkestra\ss e 85, 22607 Hamburg, Germany}\\[5mm]
\end{center}

\vspace{1cm}

\begin{abstract}
 The well--known Giudice--Masiero mechanism explains the presence of a $\mu$
 term of the order of the gravitino mass, but does not explain why the
 holomorphic mass term is absent in the superpotential. We discuss anomaly--free
 discrete symmetries which are both compatible with SU(5) unification of matter
 and the Giudice--Masiero mechanism, i.e.\ forbid the $\mu$ term in the
 superpotential while allowing the necessary K\"ahler potential term. We find
 that these are $\Z{M}^R$ symmetries with the following properties: (i) $M$ is a
 multiple of four; (ii) the Higgs bilinear $\hu\,\hd$ transforms
 trivially; (iii) the superspace coordinate $\theta$ has charge $M/4$ and,
 accordingly, the superpotential has charge $M/2$; (iv) dimension five proton
 decay operators are automatically absent. All $\Z{M}^R$ symmetries are
 anomaly--free due to a non--trivial transformation of a Green--Schwarz axion,
 and, as a consequence, a holomorphic $\mu$ term appears at the
 non--perturbative level. There is a unique symmetry that is consistent with the
 Weinberg operator while there is a class of $\Z{M}^R$ symmetries which explain
 suppressed Dirac neutrino masses.
\end{abstract}

\end{titlepage}

\section{Motivation}

The minimal supersymmetric standard model (MSSM) is a very appealing extension
of the standard model of particle physics. Supersymmetry promises to stabilize
the electroweak scale against radiative corrections.  The structure of matter
hints at unification, and the attractive picture of precision gauge unification
\cite{Dimopoulos:1981yj} enabled by supersymmetry introduces the scale of grand
unification $M_\mathrm{GUT}=\text{a few}\times10^{16}\,\mathrm{GeV}$. The MSSM
also provides a compelling dark matter candidate.

On the other hand, the MSSM  has various problems. Usually the MSSM comes with
matter or $R$ parity \cite{Farrar:1978xj,Dimopoulos:1981dw} which eliminates the
most troublesome baryon number violating interactions, and ensures the stability
of the aforementioned dark matter particle. Yet, even after imposing matter
parity, there are certain serious shortcomings. One of them is the so--called
``$\mu$ problem'' which consists in the question why the holomorphic mass term
for the Higgs bilinear is of the order of the electroweak scale.  In addition,
there is the dimension five proton decay problem
\cite{Sakai:1981pk,Weinberg:1981wj,Dimopoulos:1981dw} (cf.\ also
\cite{Hinchliffe:1992ad}).

It is hence clear that the MSSM requires additional ingredients beyond matter
parity. In this study we analyze anomaly--free discrete symmetries which forbid
the $\mu$ term. As we shall demonstrate, requiring that the symmetries be
compatible with the Giudice--Masiero solution \cite{Giudice:1988yz} to the $\mu$
problem and \SU5 leads to very restricted classes of solutions, depending on
whether neutrinos are Majorana or Dirac particles.  In the first case, the
solution is unique and even compatible with \SO{10} while in the second case the
smallness of the Dirac neutrino Yukawa coupling can be related to the
suppression of the $\mu$ parameter.

\section{Naturally suppressed $\boldsymbol{\mu}$ term and Dirac neutrino Yukawa
couplings from anomaly--free symmetries}

We start by reviewing the explanations of a suppressed $\mu$ term through
K\"ahler  potential terms in Section~\ref{sec:GiudiceMasiero}. Next, we discuss 
anomaly constraints in Section~\ref{sec:AnomalyConstraints}. In 
\Secref{sec:Proton} we comment on proton decay operators and study settings 
with SO(10) relations in \Secref{sec:SO10}. Then, we discuss the appearance of 
a suppressed holomorphic $\mu$ term and Dirac Yukawa couplings in 
\Secref{sec:NonPerturbativeMuTerm} and \Secref{sec:DiracNeutrinoMasses}, 
respectively, and give a short recap in \Secref{sec:Discussion}.

\subsection{Giudice--Masiero mechanism}
\label{sec:GiudiceMasiero}

The famous Giudice--Masiero mechanism \cite{Giudice:1988yz} provides a  solution
to the $\mu$ problem in the MSSM. Giudice and Masiero pointed out that in
supergravity an effective holomorphic $\hu\,\hd$ bilinear, i.e.\ an effective
$\mu$ term, can arise from the (non--holomorphic) K\"ahler potential term
\begin{equation}
 K~\supset~k_{\hu\hd}\,\frac{X^\dagger}{M_\mathrm{P}}\,\hu\,\hd
 +\text{h.c.}\;.
\end{equation}
Here $X$ is the (spurion) field that breaks supersymmetry and $k_{\hu\hd}$ and
$M_\mathrm{P}$ denote a coefficient and the Planck scale, respectively.
Inserting the $F$ term vacuum expectation value (VEV) $F_X$ of $X$ leads to an
effective superpotential term
\begin{equation}\label{eq:GiudiceMasiero}
 \mathscr{W}_\mathrm{eff}~\sim~\frac{F_X}{M_\mathrm{P}}\,\hu\,\hd
 ~=:~\mu_\mathrm{eff}\,\hu\,\hd\;,
\end{equation}
with $\mu_\mathrm{eff}$ of the order of the gravitino mass $m_{3/2}$, which sets
the size of soft superpartner masses in gravity mediation. 

However, for the Giudice--Masiero mechanism to work, the holomorphic
superpotential term $\mu\,\hu\,\hd$  needs to be absent in the first place, or
better forbidden by a symmetry. As it turns out, symmetries that can forbid the
$\mu$ term are rather constrained. It has been shown~\cite{Lee:2011dy} that, if
one requires the symmetry to be anomaly--free and to commute with \SU5 (in the
matter sector), it has to be an $R$ symmetry (cf.\ the similar discussion in
\cite{Hall:2002up}). As shown by Chamseddine and Dreiner
\cite{Chamseddine:1995gb}, in the MSSM gauged anomaly--free continuous $R$
symmetries are not available. On the other hand, there are strong arguments
against global symmetries (cf.\ \cite{Banks:2010zn} for a recent discussion). We
are hence led to the conclusion that the symmetry needs to be discrete. In what
follows, we therefore will only consider anomaly--free discrete $R$ symmetries. 
Specifically, we will look at one particular generator which forbids the $\mu$ 
term. This generator will generate an Abelian discrete $R$ symmetry of order
$M$,  i.e.\ a $\Z{M}^R$ symmetry.

One can actually narrow down the potential symmetries even further.  Suppose we
seek to generate an effective $\mu$ term,  \Eqref{eq:GiudiceMasiero}, from the
K\"ahler potential term.  Here we assume that $X$ is the field that breaks
supersymmetry and generates gaugino masses. Then its $F$ component has to have
minus the $R$ charge of the superpotential. One way to see this is by recalling
that gaugino masses get induced by the operator
$\int\!\D^2\theta\,X\,W_\alpha\,W^\alpha$ (with $\theta$ and $W_\alpha$ denoting
the superspace coordinate and the gauge multiplets, respectively). Since the
superpotential $R$ charge $q_\mathscr{W}$ equals twice the $R$ charge of
$\theta$, $q_\theta$, and the lowest components of $W_\alpha$  (i.e.\ the
gauginos) carry $R$ charge $q_\theta$, the $X$ superfield needs to be inert
under the (discrete) $R$ symmetry. Therefore, the Higgs bilinear $\hu\,\hd$
needs to be neutral as well. Altogether we have found that an anomaly--free and
\SU5 compatible symmetry that forbids the $\mu$ term in the MSSM has to be
discrete,  and under this symmetry,
\begin{subequations}
\begin{eqnarray}
 \theta & \to & \mathrm{e}^{2\pi\,\I\,\frac{q_\theta}{M}}\,\theta\;,\\
 \mathscr{W} & \to & \mathrm{e}^{2\pi\,\I\,\frac{q_\mathscr{W}}{M}}\,\mathscr{W}
 \qquad\text{where}~q_\mathscr{W}~=~2\,q_\theta\;,\\
 X & \to & X\;,\\
 \hu\,\hd & \to & \hu\,\hd\;.\label{eq:HuHdTrivial}
\end{eqnarray}
\end{subequations}
Here and throughout this study we normalize the discrete charges to be integer,
i.e.\ $q_\theta\in\mathbbm{Z}$.

It is immediately clear that such a symmetry allows effective superpotential
terms of the form
\begin{equation}
 \mathscr{W}~\supset~c_\Omega\,\frac{\Omega}{M_\mathrm{P}^2}\,\hu\,\hd\;,
\end{equation}
where $\Omega$ (with $R$ charge $q_\mathscr{W}$) denotes the superpotential of
some `hidden sector'. As usual, a non--trivial VEV of $\Omega$ is required to
cancel the vacuum energy. This VEV will break the $R$ symmetry, but the breaking
is hierarchically small, i.e.\ of the order of the gravitino mass $m_{3/2}$
(cf.\ the discussion in \cite{Brummer:2010fr}). That means that, apart from the
Giudice--Masiero contribution, one would expect to have a holomorphic
(`Kim--Nilles type' \cite{Kim:1983dt}) contribution to the $\mu$ parameter of
the right size.

\subsection{Anomaly constraints}
\label{sec:AnomalyConstraints}

Up to now we have only used the fact derived in \cite{Lee:2011dy} that 
\SU5--compatible and anomaly--free non--$R$ symmetries cannot forbid  the $\mu$
term. Now we discuss anomaly constraints on $\Z{M}^R$ symmetries.  These
constraints have been re--derived recently in \cite{Lee:2011dy}. However, there
only the special case $q_\theta=1$ has been considered, which is too strong a
requirement. To see this, consider a $\Z8^R$ symmetry, for which there are two
different non--trivial possibilities for the superspace charge, $q_\theta=1$ and
$q_\theta=2$. At first glance, one may think that one may rewrite the
$q_\theta=2$ case as a $\Z4^R\times\Z2$ symmetry. This is not the case since 2
and 8 are not coprime.\footnote{A simple way of seeing this is to recall that
all elements of $\Z4^R\times\Z2$ have the property that taking them to the
fourth power yields identity, which is obviously not the case for $\Z8^R$.}  The
generalization of the anomaly coefficients to arbitrary $q_\theta$ is
straightforward and deferred to appendix~\ref{sec:AnomalyCoefficients}.

After summarizing  the relevant anomaly coefficients for the MSSM in
\Secref{sec:MSSMAnomalyCoefficients} we explain in
\Secref{sec:AnomalyUniversality} why `anomaly universality' must  be imposed in
models in which the SM gauge group is unified into a simple gauge group. We then
proceed by verifying the consistency with anomaly matching in
\Secref{sec:AnomalyMatching} and show that only $R$ symmetries can forbid the
$\mu$ term in \Secref{sec:OnlyR}. Finally, we derive constraints on the  order
$M$ in \Secref{sec:Mdividesfour} and comment on the (ir)relevance of the
universality of the mixed hypercharge anomaly in \Secref{sec:A1notneeded}.

\subsubsection{$\boldsymbol{\Z{M}^R}$ anomaly coefficients in the MSSM}
\label{sec:MSSMAnomalyCoefficients}

In the case of the MSSM the anomaly coefficients 
$A_3^R:=A_{\SU3_\mathrm{C}-\SU3_\mathrm{C}-\Z{M}^R}$,
$A_2^R:=A_{\SU2_\mathrm{L}-\SU2_\mathrm{L}-\Z{M}^R}$ and
$A_1^R:=A_{\U1_Y-\U1_Y-\Z{M}^R}$  read 
\begin{subequations}\label{eq:AnomalyCoefficientsMSSM}
\begin{eqnarray}
 A_3^R
 & = &
 \frac{1}{2}\sum_{g=1}^3
 \left(3q_{\boldsymbol{10}}^g+q_{\overline{\boldsymbol{5}}}^g\right)
 -3q_\theta
 \;,\\
 A_2^R
 & = &
 \frac{1}{2}\sum_{g=1}^3
 \left(3q_{\boldsymbol{10}}^g+q_{\overline{\boldsymbol{5}}}^g\right)
 +\frac{1}{2}\left(\qhu+\qhd\right)-5q_\theta
 \;,\\
 A_1^R
 & = &
 \frac{1}{2}\sum_{g=1}^3
 \left(3q_{\boldsymbol{10}}^g+q_{\overline{\boldsymbol{5}}}^g\right)
 +\frac{3}{5}\left[\frac{1}{2}\left(\qhu+\qhd\right)-11q_\theta\right]
 \;.
\end{eqnarray}
\end{subequations}
Here, $q_{\boldsymbol{10}}^g$ and $q_{\boldsymbol{\overline{5}}}^g$ denote the
\SU5--universal $R$ charges of the MSSM superfields
$(Q^g,\overline{U}^g,\overline{E}^g)$ and $(\overline{D}^g,L^g)$, respectively,
and $g$ represents the flavor index. Accordingly, matter fermions and Higgsinos
have charges $q-q_\theta$ while gauginos have charge $q_\theta$.

\subsubsection{Anomaly universality and discrete Green--Schwarz mechanism}
\label{sec:AnomalyUniversality}

If the standard model gauge group is to be unified into \SU5 or \SO{10}, a
necessary condition for anomaly cancellation is the universality (cf.\ the
discussion in appendix~\ref{app:AnomalyUniversality})
\begin{equation}
 A_3^R~=~A_2^R~=~A_1^R~=~\rho\mod \eta\;.
 \label{eq:anomaly-universality-R}
\end{equation}
Here we introduce
\begin{equation}\label{eq:eta}
 \eta~:=~\left\{\begin{array}{ll}
 M/2\;, & \text{if $M$ even}\;,\\
 M\;, & \text{if $M$ odd}\;.
 \end{array}\right.
\end{equation}
$\rho$ is a constant which indicates whether or not a Green--Schwarz (GS)
mechanism \cite{Green:1984sg} is at work. Specifically, $\rho$ is related to the
discrete shift of the GS axion  (see \Eqref{eq:DiscreteShiftDilaton} in
appendix~\ref{app:DiscreteGS}).  $\rho=0$ means that the symmetry is
anomaly--free in the conventional sense, i.e.\ without GS mechanism.

At this point, we would like to comment on certain important properties of the
Green--Schwarz mechanism and its discrete version as there seems to be some
confusion in the literature:
\begin{enumerate}
 \item Although the GS mechanism plays a prominent role in string theory, it
  does not rely on strings. In fact, as shown in
  appendix~\ref{app:DiscreteGS}, it can entirely be understood in (the
  path integral formulation of) quantum field theory.
 \item Unlike in the continuous case, for discrete symmetries the transformation
  of the axion is only fixed modulo $\eta$. It will be interesting to see whether
  this ambiguity can be fixed somehow, e.g.\ in explicit string--derived models.
 \item In the continuous case, the axion has to be massless for the shift
 symmetry to be a symmetry of the Lagrangean. That is, the axion potential needs
 to be flat. By contrast, in the discrete case the potential is only required to
 be periodic, i.e.\ invariant under the discrete shift,  
 \Eqref{eq:DiscreteShiftDilaton}. Therefore the axion may have a non--trivial
 mass prior to the breakdown of the symmetry. This is, in a way, somewhat
 surprising as it means that a massive (and bosonic) state can contribute to an
 anomaly. Of course, in both cases the symmetry will be broken (spontaneously)
 once the axion $a$ acquires its VEV.
\end{enumerate}

\subsubsection{Anomaly matching}
\label{sec:AnomalyMatching}

It is instructive to use 't Hooft anomaly matching \cite{'tHooft:1980xb} (see
\cite{Csaki:1997aw} for discrete anomaly matching) in order to constrain the
properties of anomaly--free GUT--compatible $\Z{M}^R$ symmetries. At the \SU5
level, there is only one anomaly coefficient $A_{\SU5^2-\Z{M}^R}$, which we can
split into three parts,
\begin{equation}\label{eq:AnomalyMatching1}
 A_{\SU5^2-\Z{M}^R}
 ~=~
 A_{\SU5^2-\Z{M}^R}^\mathrm{matter}
 +
 A_{\SU5^2-\Z{M}^R}^\mathrm{extra}
 +5q_\theta
 \;.
\end{equation}
The first term contains the contribution of matter and is given by
\begin{equation}
 A_{\SU5^2-\Z{M}^R}^\mathrm{matter} 
 ~=~ 
 \frac{1}{2}\sum_{g=1}^3
 \left(3q_{\boldsymbol{10}}^g+q_{\overline{\boldsymbol{5}}}^g\right)
 -6q_\theta\;.
\end{equation}
Here, we used \Eqref{eq:AGGZN} with Dynkin indices $\ell(\crep{5}) =
\frac{1}{3}\ell(\rep{10}) =\frac{1}{2}$. The second term in
\eqref{eq:AnomalyMatching1}, $A_{\SU5^2-\Z{M}^R}^\mathrm{extra}$,  denotes the
contributions of additional fields, e.g.\ the SM and \SU5 breaking  Higgs.
Finally, the last term in \eqref{eq:AnomalyMatching1} represents the  gaugino
contribution for \SU5. Yet, by considering the $\SU3_\mathrm{C}$ and
$\SU2_\mathrm{L}$ subgroups  of \SU5, one can introduce two anomaly
coefficients  $A^{\SU5}_{{\SU3_\mathrm{C}^2}-\Z{M}^R} =
A^{\SU5}_{{\SU2_\mathrm{L}^2}-\Z{M}^R}$ at the GUT level, 
\begin{subequations}\label{eq:AnomalyMatching2}
\begin{eqnarray}
 A^{\SU5}_{{\SU3_\mathrm{C}^2}-\Z{M}^R}
 & = &
 A_{\SU3_\mathrm{C}^2-\Z{M}^R}^\mathrm{matter}
 +
 A_{\SU3_\mathrm{C}^2-\Z{M}^R}^\mathrm{extra}
 +3q_\theta+\frac{1}{2}\cdot2\cdot2\cdot q_\theta
 \;,
 \\
 A^{\SU5}_{{\SU2_\mathrm{L}^2}-\Z{M}^R}
 & = &
 A_{\SU2_\mathrm{L}^2-\Z{M}^R}^\mathrm{matter}
 +
 A_{\SU2_\mathrm{L}^2-\Z{M}^R}^\mathrm{extra}
 +2q_\theta+\frac{1}{2}\cdot2\cdot3\cdot q_\theta
 \;,
\end{eqnarray}
\end{subequations}
where we artificially split the gaugino contributions into those from the
adjoint representations of $\SU2_\mathrm{L}$ or $\SU3_\mathrm{C}$, respectively,
and in those coming from the extra gauginos in the
$(\rep{3},\rep{2})_{-\nicefrac{5}{6}}\oplus(\crep{3},\rep{2})_{\nicefrac{5}{6}}$
representation. Assume now there is some (unspecified) mechanism that breaks the
GUT symmetry down to the SM symmetry, and thus removes the extra gauginos, while
leaving $\Z{M}^R$ unbroken.\footnote{If one is to obtain the exact MSSM spectrum
after GUT breaking, this mechanism cannot be spontaneous symmetry breaking in
four dimensions \cite{Fallbacher:2011xg}. On the other hand, extra dimensions, 
especially in the framework of heterotic orbifolds, naturally can give discrete 
$R$ symmetries as remnants of higher dimensional Lorentz symmetry, see e.g.\ 
\cite{Nilles:2012cy}.} Then, the coefficients
\begin{subequations}\label{eq:AnomalyMatching3}
\begin{eqnarray}
A^{\SU5\text{ broken}}_{{\SU3_\mathrm{C}^2}-\Z{M}^R} & = & A^{\SU5}_{{\SU3_\mathrm{C}^2}-\Z{M}^R} - 2q_\theta \;,\\
A^{\SU5\text{ broken}}_{{\SU2_\mathrm{L}^2}-\Z{M}^R} & = & A^{\SU5}_{{\SU2_\mathrm{L}^2}-\Z{M}^R} - 3q_\theta
\end{eqnarray}
\end{subequations}
cannot be equal, i.e.\ the anomaly coefficients cannot be universal, unless 
there are split multiplets contributing to 
$A_{{\SU{N}^2}-\Z{M}^R}^\mathrm{extra}$  (where we use  
$A_{\SU3_\mathrm{C}^2-\Z{M}^R}^\mathrm{matter}=A_{\SU2_\mathrm{L}^2-\Z{M}^R}^\mathrm{matter}$). 
That is, \emph{'t Hooft anomaly matching for (discrete) $R$  symmetries implies
the presence of split multiplets below the GUT scale}. 

\subsubsection{Only $\boldsymbol{R}$ symmetries can forbid the
$\boldsymbol{\mu}$ term}
\label{sec:OnlyR}

Given that SM matter furnishes complete \SU5 representations and the attractive
picture of MSSM gauge unification, arguably the most plausible candidates for
such split multiplets are the Higgs fields. Requiring that the Higgs fields
cancel the mismatch of gaugino contributions to the anomalies, we obtain 
\begin{equation}
\frac{1}{2}\left(q_{\hu}+q_{\hd}-2q_\theta\right) ~=~ q_\theta \mod \eta \;,
\end{equation}
implying
\begin{subequations}\label{eq:qHuqHd1}
\begin{eqnarray}
q_{\hu}+q_{\hd} & = & 4q_\theta\mod2\eta\\
\label{eq:AnomalyConditionOnHiggsCharge}
& = & 2q_\mathscr{W}\mod2\eta \\
& \neq & q_\mathscr{W}\mod M \quad \text{for}~q_\mathscr{W}~\neq~0\mod M\;.
\end{eqnarray}
\end{subequations}
Therefore, non--$R$ symmetries with $q_\theta = q_\mathscr{W} = 0$ cannot 
forbid the $\mu$ term. But in case of non--trivial $\Z{M}^R$ symmetries (i.e.\ 
$M\ge3$) the $\mu$ term will always be forbidden, as it should be, since  only
chiral contributions can `repair' the gaugino mismatch.

A remark is in order to show that $\Z{M}^R$ with $q_\mathscr{W}=0\mod M$  is not
an $R$ symmetry. In the case $q_\mathscr{W}=0\mod M$ we find two  solutions for
$q_\theta$: either $q_\theta=0$, such that the symmetry is  clearly non--$R$, or
(for $M$ even) $q_\theta=M/2$. However, since the  transformation $\theta
\mapsto -\theta$ and $\Psi \mapsto -\Psi$ for all  fermions $\Psi$ is always a
symmetry, one can shift the $\Z{M}^R$ charges by  $M/2$ such that again
$q_\theta=0$. Hence, $\Z{M}^R$ with  $q_\mathscr{W}=0\mod M$ is equivalent to a
non--$R$ symmetry \cite{Dine:2009swa}.

\subsubsection{Constraints on the order $\boldsymbol{M}$}
\label{sec:Mdividesfour}

Using \Eqref{eq:qHuqHd1} and assuming a Giudice--Masiero--like mechanism such
that $q_{\hu}+q_{\hd}=0 \mod M$ from \eqref{eq:HuHdTrivial}, we obtain
\begin{equation}\label{eq:2qW=0}
 2q_\mathscr{W}~=~0\mod M\;,
\end{equation}
which implies, given the freedom to choose $q_\mathscr{W}$ between $0$ and
$M-1$, that the only non--trivial solution for even $M$ is $q_\mathscr{W}=M/2$.
For odd $M$ there is no non--trivial solution. Since the superpotential charge 
is given by $q_\mathscr{W}=2q_\theta$, the order $M$ has to be divisible by 
$4$. Hence we can focus on 
\begin{equation}
 M~=~4\times\text{integer}\quad\text{and}\quad q_\theta~=~M/4
\end{equation}
in the rest of our discussion.

\subsubsection{No additional condition from $\boldsymbol{A_1^R}$}
\label{sec:A1notneeded}

Subtracting $A_3^R$ from $A_1^R$ yields
\begin{equation}
 \frac{3}{5}\left[\frac{1}{2}\left(q_{\hu}+q_{\hd}\right)-11q_\theta\right]
 +3q_\theta
 ~=~0\mod\eta\;.
\end{equation}
Since $M$ is even and $q_{\hu}+q_{\hd}=0\mod M$ by \eqref{eq:HuHdTrivial}, this
equation is equivalent to
\begin{equation}
 3\,k\,M+2\,(15-33)\,q_\theta~=~ 5\ell\,M
\end{equation}
with some integers $k$ and $\ell$. That is,
\begin{equation}\label{eq:Redundant}
 36\,q_\theta~=~\left[3\,k-5\,\ell\right]\,M~=~\mathbbm{Z}\cdot M\;.
\end{equation}
For a given order $M$, this relation constrains $q_\theta$.  However, we know
already from our discussion below \Eqref{eq:2qW=0} that $M$ needs to be an
integral multiple of 4,  such that \eqref{eq:Redundant} does not lead to
an additional constraint.

\subsection{Family--independent symmetries and proton decay}
\label{sec:Proton}

Let us now assume further that the discrete symmetry be Abelian, i.e.\ of
$\Z{M}^R$ type, with family--independent charges. Assuming the presence of 
Yukawa couplings, symmetries with the above properties have automatically the 
virtue of solving the dimension five proton decay problem of the MSSM, as we 
will see in the following. 

The requirement that up- and down--type Yukawa couplings be allowed,
\begin{subequations}
\begin{eqnarray}
 2q_{\rep{10}}+q_{\hu} & = & q_\mathscr{W}\mod M\;,
 \label{eq:YuAllowed}\\
 q_{\rep{10}}+q_{\crep{5}}+q_{\hd} & = & q_\mathscr{W}\mod M\;,
 \label{eq:YdAllowed}
\end{eqnarray}
\end{subequations}
implies
\begin{equation}\label{eq:YukawasAllowed}
 3q_{\boldsymbol{10}}+q_{\boldsymbol{\overline{5}}}+\qhu+\qhd
 ~=~
 2q_\mathscr{W}\mod M\;.
\end{equation}
Imposing \eqref{eq:HuHdTrivial} gives
\begin{equation}\label{eq:dim5automatic}
 3q_{\boldsymbol{10}}+q_{\boldsymbol{\overline{5}}}
 ~=~
 2q_\mathscr{W}\mod M ~\neq~ q_\mathscr{W}\mod M\;,
\end{equation}
for an $R$ symmetry (i.e.\ for $q_\mathscr{W}\neq 0\mod M$), showing that the 
troublesome dimension five operators 
$\boldsymbol{10}\,\boldsymbol{10}\,\boldsymbol{10}\,\overline{\boldsymbol{5}}$
are automatically forbidden whenever the Yukawa couplings are
allowed. In \cite{Lee:2011dy} the same conclusion was obtained from anomaly
cancellation.

Recalling further that $2q_\mathscr{W}=4q_\theta=M$ leads us to the conclusion 
that 
\begin{equation}
 q_{\crep{5}}~=~-3q_{\rep{10}}\mod M\;.
\end{equation}
This means that the contributions of matter fields to the anomaly coefficients
\eqref{eq:AnomalyCoefficientsMSSM} vanish, and that the universal anomaly
coefficients are simply given by 
\begin{equation}\label{eq:rho=qtheta}
 A_i^R~=~\rho~=~q_\theta \mod M/2
\end{equation}
for $1\le i\le 3$.

Next, one can also discuss proton decay originating from the dimension four
operator
$\boldsymbol{10}\,\overline{\boldsymbol{5}}\,\overline{\boldsymbol{5}}$. This
operator has $R$ charge
\begin{equation}
 q_{\rep{10}} + 2q_{\crep{5}}~=~-5q_{\rep{10}} \mod M\;.
\end{equation}
Hence, these operators are also forbidden if $-5q_{\rep{10}} \neq 
q_\mathscr{W}\mod M$, or equivalently $10q_{\rep{10}} \neq k\,M$ with $k$ odd.

\subsection{Imposing SO(10) relations}
\label{sec:SO10}

Let us now comment on the special case that the $\Z{M}^R$ symmetry commutes with
\SO{10} for the matter fields, i.e.\ $q_{\rep{10}}=q_{\crep{5}}=q_{\rep{16}}$. 
The requirements that the up--type and down--type quark Yukawa couplings be 
allowed imply that $q_{\hu}=q_{\hd}=:q_H$ ($\text{mod}\, M$). Furthermore, from
the  anomaly universality condition (\ref{eq:AnomalyConditionOnHiggsCharge}) we
find  $q_H = q_\mathscr{W}\mod\eta$. In the following, we consider two cases: in
case  (i) we demand in addition the Weinberg neutrino mass operator, and in case
(ii)  a Giudice--Masiero--like mechanism.

(i) If we require the Weinberg neutrino mass operator, i.e.\ $2q_{\rep{16}} + 2q_H = 
q_\mathscr{W}\mod M$, we find $M=4m$, $m\in\mathbbm{N}$ and
\begin{equation}
q_{\theta}~=~m \;,\; q_\mathscr{W}=2m \;,\; q_H = 0\;\text{ and }\; q_{\rep{16}}=m\;.
\end{equation}
This symmetry automatically allows for the Giudice--Masiero term and the 
universal anomaly coefficients $A_i^R = m \neq 0$ indicate a discrete GS 
mechanism. The simplest case $m=1$ is the $\Z4^R$ symmetry discussed in 
\cite{Babu:2002tx,Lee:2010gv}. All other cases are just trivial extensions as
long as one considers the MSSM states only. Of course, if additional states are
introduced, they can have $\Z{4m}^R$ charges in such a way that one cannot
reduce it to $\Z4^R$. Another version of the uniqueness  proof of $\Z4^R$ can be
found in \cite{Lee:2011dy}. However, the analysis in  \cite{Lee:2011dy} assumed
that $q_\theta=1$. Here we show that uniqueness also  survives the
generalization to general $q_\theta\ne1$.

(ii) If we do not require the Weinberg neutrino mass operator but a 
Giudice--Masiero--like mechanism, i.e.\ $2q_H = 0\mod M$, there are two cases:
both  cases have $M=4m$, $m\in\mathbbm{N}$, $q_{\theta} = m$ and
$q_\mathscr{W}=2m$. In addition, in the first case we get $q_H = 0$ as discussed
above in case (i),  and in the second one we find $q_H=M/2=2m$ and
$q_{\rep{16}}=2\ell\, m$ with  $\ell\in\Z{}$. However, this choice forbids the
Weinberg neutrino mass operator.

\subsection{Non--perturbative holomorphic $\boldsymbol{\mu}$ term}
\label{sec:NonPerturbativeMuTerm}

If the above discrete $R$ symmetry appears anomalous, i.e.\ if anomaly freedom
is due to a GS mechanism (see appendix~\ref{app:GS} for a discussion of its
discrete variant), then such holomorphic contributions will appear as arising at
the non--perturbative level \cite{Lee:2010gv,Lee:2011dy}. To see this, recall
that the superfield $S$ containing the axion $a$, i.e.\ $S|_{\theta=0}=s+\I\,a$,
needs to enter the gauge--kinetic function, or, in other words,
$\mathscr{L}\supset\int\!\D^2\theta\,f_S\, S\,W_\alpha W^\alpha$ (with some
coefficient $f_S$).  Non--invariant terms in the superpotential can be made
invariant by multiplying them by $\mathrm{e}^{-b\,S}$ with appropriate $b$. As
$s$ controls $1/g^2$ such terms go like $\mathrm{e}^{-b'/g^2}$, i.e.\ have the
form of instanton contributions. This then fits nicely into the scheme of
dynamical supersymmetry breaking \cite{Witten:1981nf} (see also the more recent
discussion on ``retrofitting''  \cite{Dine:2006gm}), where the scale for
supersymmetry breaking is set by a gaugino condensate \cite{Nilles:1982ik}, or a
more complicated dynamical term (see e.g.\ \cite{Intriligator:2007py} for a
review of simple models).

\subsection{Small Dirac neutrino Yukawa couplings}
\label{sec:DiracNeutrinoMasses}

By relating them to supersymmetry breaking one may explain suppressed
neutrino Dirac Yukawa couplings
\cite{ArkaniHamed:2000bq,Borzumati:2000mc,MarchRussell:2004uf}. That is,
similarly to the $\mu$ term, one can get effective Dirac neutrino Yukawa
couplings from the K\"ahler potential terms
\begin{subequations}\label{eq:YnuKaehler}
\begin{equation}\label{eq:YnuKaehler1}
 K~\supset~k_{L\hu\bar\nu}\,\frac{X^\dagger}{M_\mathrm{P}^2}\,
 L\,\hu\,\bar\nu+\text{h.c.}
\end{equation}
as well as 
\begin{equation}\label{eq:YnuKaehler2}
 K~\supset~k_{\hd^\dagger L\bar\nu}\,\frac{1}{M_\mathrm{P}}\,
 \hd^\dagger\,L\,\bar\nu+\text{h.c.}
\;.
\end{equation}
\end{subequations}
Here, in an obvious notation, $\bar\nu$ denotes the right--handed neutrino
superfield(s), $k_{L\hu\bar\nu}$ and $k_{\hd^\dagger L\bar\nu}$ are
dimensionless coefficients, and we suppress flavor indices. The first term
\eqref{eq:YnuKaehler1} leads to Dirac neutrino masses when $X$ attains its
$F$--term VEV, $\langle F_X\rangle\sim m_{3/2}\,M_\mathrm{P}$, while in the case
of \eqref{eq:YnuKaehler2} one has to observe that, due to the presence of the
`non--perturbative' $\mu$ term, also $\hd$ attains an $F$ term VEV, $\langle
F_{\hd}\rangle\sim \mu\,\langle \hu\rangle\sim m_{3/2}\,v_\mathrm{EW}$. As
$q_{\hu}+q_{\hd}=0\mod M$, both terms are allowed if 
$q_{\bar\nu}+q_{\hu}+q_L=0\mod M$, which is precisely the condition that an
effective holomorphic $Y_\nu$ term is allowed. Altogether we find, analogous to
what we have discussed around \eqref{eq:GiudiceMasiero}, that effective neutrino
Yukawa couplings 
\begin{equation}
 Y_\nu~\sim~\frac{m_{3/2}}{M_\mathrm{P}}~\sim~\frac{\mu}{M_\mathrm{P}}
\end{equation}
will arise. For $m_{3/2}$ in the multi--TeV range this can lead to realistic
Dirac neutrino masses. If we are to connect the suppression of $Y_\nu$ to the
smallness of the $\mu$ term, it is natural to assume that the neutrino Yukawa
coupling is forbidden by the same $R$ symmetry that also forbids $\mu$. As
discussed above,  $L\,\hu\,\bar\nu$ has to have $R$ charge 0. Moreover, there
will also be holomorphic contributions to the Yukawa coupling. That is, even if
both $k_{L\hu\bar\nu}$ and $k_{\hd^\dagger L\bar\nu}$ vanish, Dirac Yukawa
couplings of the order $m_{3/2}/M_\mathrm{P}$ will get induced, where, as in our
discussion of the $\mu$ term, $m_{3/2}$ represents the order parameter for $R$
symmetry breaking.

\subsection{Discussion}
\label{sec:Discussion}

We have surveyed anomaly--free symmetries which forbid the $\mu$ term and are
consistent with the Giudice--Masiero mechanism and \SU5. We find that these are
discrete $R$ symmetries $\Z{M}^R$ with $M=4m$, $m\in\mathbbm{N}$. The $R$
charges of the $\hu\,\hd$ are such that  one expects a holomorphic contribution
to the $\mu$ term of  similar size. That is, the Giudice--Masiero mechanism
strongly suggests the presence of additional holomorphic contributions to the
effective $\mu$ term!

Assuming further that the symmetries allow the up- and down--type Yukawa 
couplings and commute with flavor we find  that they automatically forbid the
troublesome dimension five proton decay  operators and in many cases those of
dimension four. Interestingly, all these symmetries require a GS axion for
anomaly  cancellation. That is, these symmetries appear to be broken at the
non--perturbative level.  In other words, imposing compatibility with the
Giudice--Masiero mechanism leads us to a situation in which a holomorphic $\mu$
term appears at the non--perturbative level, i.e.\ in a way the Giudice--Masiero
term is unnecessary.

\section{Classification and models}
\label{sec:Classification}

In this section, we explore anomaly--free discrete symmetries that solve
some of the most severe problems of the MSSM. We will demand that the symmetry
\begin{enumerate}
 \item is flavor--universal and Abelian, i.e.\ a $\Z{M}^R$ symmetry;\label{cond1}
 \item commutes with \SU5;
 \item forbids the $\mu$ term perturbatively;
 \item allows the usual Yukawa couplings;\label{cond5}
\end{enumerate}
After revisiting in Section~\ref{sec:Majorana} the scan performed in \cite{Lee:2011dy},
where Majorana neutrinos were considered,
we turn to the Dirac case in Section~\ref{sec:Dirac}.

\subsection{Models with Majorana neutrinos}
\label{sec:Majorana}

In \cite{Lee:2010gv,Lee:2011dy}, anomaly--free discrete $R$ symmetries 
with $q_\theta=1$ were studied which satisfy the requirements 
\ref{cond1}--\ref{cond5} and in addition
\begin{enumerate}\setcounter{enumi}{4}
 \item allow the Weinberg neutrino mass operator.
\end{enumerate}
It was found that there are only five phenomenologically attractive symmetries
that commute with \SU5, one of which, a simple $\Z4^R$ symmetry, commutes also
with \SO{10}. Further, the $\mu$ term, while perturbatively forbidden, appears
at the non--perturbative level in four out of the five symmetries, and thus can
explain its suppression. There is one symmetry which is anomaly--free without GS
contribution; here anomaly freedom requires the number of generations to be a
multiple of 3 \cite{Evans:2011mf} (for a similar connection between the number
of generations and anomaly--free non--$R$ symmetries see
\cite{Hinchliffe:1992ad,Mohapatra:2007vd}).

In the classification of \cite{Lee:2011dy}, $\Z4^R$ appears to be particularly
attractive. Apart from the fact that it is the unique solution that commutes
with \SO{10}, only $\Z4^R$ provides a real solution to the $\mu$ problem. 
In this case,  the discrete charges of $H_u$ and $H_d$ add up to $0\mod M=4$
such that the $\mu$ parameter will be of the order of the gravitino mass, i.e.\
the order parameter of $R$ breaking. This feature is not shared by the other
four $\Z{M}^R$ symmetries, as also can be seen from our analysis in
\Secref{sec:SO10}. In particular, it was argued that $\mu\sim m_{3/2}$ for the
case of $\Z{4}^R$.  To substantiate these claims, an explicit string model with
exact MSSM spectrum and the $\Z4^R$ symmetry was constructed in which the
relation $\mu\sim \langle\mathscr{W}\rangle\sim m_{3/2}$ is due to gauge
invariance in extra dimensions \cite{Kappl:2010yu}.

Assuming in addition a Giudice--Masiero--like mechanism, one can see that 
$\Z4^R$ is the unique solution also for general $q_\theta$ as follows.  From the
requirement that the Weinberg operator be allowed we infer that
\begin{equation}\label{eq:WeinbergAllowed}
 2q_{\crep{5}}+2q_{\hu}~=~2q_\theta\mod M
 \quad\curvearrowright\quad
 q_{\crep{5}}~=~q_\theta-q_{\hu}\mod M/2\;.
\end{equation}
On the other hand, from the down--type Yukawa coupling it follows
\begin{equation}\label{eq:Yd}
 q_{\rep{10}}~=~-q_{\crep{5}}-q_{\hd}+2q_\theta\mod M
 ~\stackrel{(\ref{eq:WeinbergAllowed})}{=}~
 q_\theta+q_{\hu}-q_{\hd}\mod M/2\;.
\end{equation}
Demanding that the up--type Yukawa coupling be allowed leads to
\begin{eqnarray}
 q_{\hu}& = & 2q_\theta-2q_{\rep{10}}\mod M
 \nonumber\\
 & \stackrel{(\ref{eq:Yd})}{=} &
 -2q_{\hu}+2q_{\hd}\mod M~=~-4q_{\hu}\mod M\;,
\end{eqnarray}
such that $5q_{\hu}=0\mod M$. This means that $q_{\hu}=0\mod M$ unless the order
is a multiple of 5. In the latter case we can write the $\Z{M}^R$ symmetry as
$\Z5\times\Z{M/5}^R$ where the $\Z5$ factor is a non--$R$ symmetry. Hence we can
focus on $q_{\hu}=0\mod M$, which implies, by \eqref{eq:HuHdTrivial}, that
$q_{\hd}=0\mod M$. Then Equations~\eqref{eq:WeinbergAllowed} and \eqref{eq:Yd}
imply
\begin{equation}
 q_{\rep{10}}~=~q_{\crep{5}}~=~q_{\theta} \mod M\;.
\end{equation}
That is, the symmetry commutes with \SO{10} in the matter sector.
We already know from our discussion in Section~\ref{sec:SO10} that the only
meaningful $R$ symmetry with this property is $\Z4^R$.

We also scanned the discrete $\Z{M}^R$ symmetries up to order 200 with  general
$q_\theta$ without assuming a Giudice--Masiero--like mechanism. We  obtain,
apart from the symmetries of Tables 2.1 and 2.2 of \cite{Lee:2011dy},  only a
few new symmetries. However, as we show in the following in an example,  these
additional symmetries are redundant: consider a $\Z{20}^R$ symmetry with 
$(q_{\rep{10}},q_{\crep{5}},q_{\hu},q_{\hd},q_{\theta})=\left(1,17,8,52,5\right)$.
This is equivalent to a $\Z4^R\times\Z5$ symmetry with charge assignment
$\left((1,3),(1, 1), (0, 4), (0, 1), (1, 0)\right)$. The \Z5 is nothing but the
non--trivial center of \SU5, i.e.\ it does not forbid any couplings (see the
discussion in \cite{Csaki:1997aw,Petersen:2009ip}) and the (non--trivial) 
$\Z4^R$ factor is the one just discussed in the last paragraph.

\subsection{Models with Dirac neutrinos}
\label{sec:Dirac}

By modifying the above conditions, i.e.\ by demanding that the symmetry
\begin{enumerate}\setcounter{enumi}{4}
 \item forbids the Weinberg neutrino mass operator perturbatively 
\end{enumerate}
and
\begin{enumerate}\setcounter{enumi}{5}
 \item is compatible with the Giudice--Masiero mechanism
\end{enumerate}
we obtain further interesting discrete $R$ symmetries. Some sample symmetries
are listed in \Tabref{tab:Classification}. Anomaly--free (non--$R$) $\Z{N}$ 
symmetries which allow for Dirac neutrino Yukawa couplings have been discussed 
in \cite{Luhn:2007gq}.
\begin{table}[h]
\centerline{\subtable[$\Z{M}^R$ symmetries.]{$\displaystyle
\begin{array}{ccccccccc}
 M & q_{\boldsymbol{10}} & q_{\boldsymbol{\overline{5}}} & q_{\hu} & q_{\hd}
 & q_{\theta} & \rho & q_{\bar\nu}\\
 \hline
  4 & 0 & 0 & 2 & 2 & 1 & 1 & 2 \\
  4 & 2 & 2 & 2 & 2 & 1 & 1 & 0 \\
  8 & 1 & 5 & 2 & 6 & 2 & 2 & 1 \\
 12 & 1 & 9 & 4 & 8 & 3 & 3 & 11 \\
 12 & 2 & 6 & 2 & 10 & 3 & 3 & 4 \\
 12 & 4 & 0 & 10 & 2 & 3 & 3 & 2 \\
 16 & 1 & 13 & 6 & 10 & 4 & 4 & 13 \\
 24 & 1 & 21 & 10 & 14 & 6 & 6 & 17 \\
 28 & 1 & 25 & 12 & 16 & 7 & 7 & 19 \\
 28 & 2 & 22 & 10 & 18 & 7 & 7 & 24 \\
 28 & 4 & 16 & 6 & 22 & 7 & 7 & 6 \\
 32 & 1 & 29 & 14 & 18 & 8 & 8 & 21 \\
 36 & 1 & 33 & 16 & 20 & 9 & 9 & 23 \\
 36 & 2 & 30 & 14 & 22 & 9 & 9 & 28 \\
 36 & 4 & 24 & 10 & 26 & 9 & 9 & 2 
\end{array}
$}
\qquad
\subtable[Residual symmetries.]{$\displaystyle
\begin{array}{cccccc}
 M' & q_{\rep{10}} & q_{\crep{5}} & q_{\hu} & q_{\hd} & q_{\bar\nu} \\
 \hline
 2 & 0 & 0 & 0 & 0 & 0 \\
 2 & 0 & 0 & 0 & 0 & 0 \\
 4 & 1 & 1 & 2 & 2 & 1 \\
 6 & 1 & 3 & 4 & 2 & 5 \\
 3 & 1 & 0 & 1 & 2 & 2 \\
 3 & 2 & 0 & 2 & 1 & 1 \\
 8 & 1 & 5 & 6 & 2 & 5 \\
 12 & 1 & 9 & 10 & 2 & 5 \\
 14 & 1 & 11 & 12 & 2 & 5 \\
 7 & 1 & 4 & 5 & 2 & 5 \\
 7 & 2 & 1 & 3 & 4 & 3 \\
 16 & 1 & 13 & 14 & 2 & 5 \\
 18 & 1 & 15 & 16 & 2 & 5 \\
 9 & 1 & 6 & 7 & 2 & 5 \\
 9 & 2 & 3 & 5 & 4 & 1 
\end{array} 
$}
}
\caption{Classification of anomaly--free discrete $R$ symmetries that forbid
neutrino masses perturbatively. We restrict to orders $\le36$. (a) shows some
sample symmetries. The equality between $q_\theta$ and $\rho$ is due to
\Eqref{eq:rho=qtheta}. The charge of the right--handed neutrino superfield
$\bar\nu$ is determined by the requirement that $q_{\bar\nu}+q_{\hu}+q_L=0\mod
M$ (cf.\ the discussion below \eqref{eq:YnuKaehler}). In (b) we display the residual symmetries that remain after the
(`hidden sector') superpotential acquires its VEV.}
\label{tab:Classification}
\end{table}
The symmetries of \Tabref{tab:Classification} are inequivalent. One way of 
verifying this is to check whether or not two given charge assignments are 
equivalent by computing their Hilbert superpotential basis 
\cite{Kappl:2011vi}. Only if the bases coincide, the
assignments are equivalent. In the case of $R$ symmetries, the Hilbert
superpotential basis comprises homogeneous and inhomogeneous elements, or
monomials.  Every possible superpotential term contains precisely one
inhomogeneous monomial and an arbitrary number of homogeneous monomials. In
appendix \ref{app:Hilbert} we list the Hilbert superpotential basis for examples
with the $\Z{12}^R$ symmetries.

\subsubsection{Comments on the $\boldsymbol{\Z8^R}$ symmetry}

One of simplest charge assignments appears to be the one of the $\Z8^R$
symmetry. Clearly the usual Yukawa couplings
$\boldsymbol{10}\,\boldsymbol{10}\,\hu$ and
$\boldsymbol{10}\,\boldsymbol{\overline{5}}\,H_d$ are allowed. Further, the
Higgs bilinear $\hu\,\hd$ has $R$ charge $0\mod8$. If we assign the
right--handed neutrino $\bar\nu$ $R$ charge 1, the Dirac neutrino Yukawa
coupling will also be induced by $R$ breaking. That is, we will have an
effective superpotential which is schematically of the form
\begin{equation}
 \mathscr{W}_\mathrm{eff}~\sim~m_{3/2}\,\hu\,\hd
 +\frac{m_{3/2}}{M_\mathrm{P}}\,L\,\hu\,\bar\nu
 +\frac{m_{3/2}}{M_\mathrm{P}^2}\,Q\,Q\,Q\,L\;.
\end{equation}
Here we suppress flavor indices. Once the superpotential of the hidden sector
acquires a VEV, the $\Z8^R$ is spontaneously broken down to a $\Z4$ symmetry
under which all matter fields have charge 1 and the Higgs fields have charge 2
(\Tabref{tab:Classification}~(b)). Of course, this symmetry gets broken down to
the usual matter (or `$R$') parity once the Higgs scalars attain their VEVs.

The Hilbert superpotential basis \cite{Kappl:2011vi} for this model (setting all
quarks to zero) is given by the inhomogeneous monomials
\begin{eqnarray}
 & & \bar\nu^4~; 
 ~\lle\,\bar\nu~;
 ~\lhde~;
 ~\lle^4~;
 ~\lle^2\,\lhu^2~;
 ~\lhu^4\;,
\end{eqnarray}
while the homogeneous monomials are
\begin{eqnarray}
& & \bar\nu^8~; 
~\lhu\,\bar\nu~;
~\lhu^8~; 
~\lle^5\,\bar\nu~; 
~\lle^4\,\lhde~; \nonumber\\
 & & 
~\hu\,\hd~;
~\lle\,\bar\nu^5~; 
~\lhde\,\bar\nu^4~; 
~\lle^2\,\lhde\,\lhu^2~; \nonumber \\
 & & 
~\lle^8~; 
~\lhde^2~; 
~\lle\,\lhde\,\bar\nu~; 
~\lle^2\,\bar\nu^2~; \nonumber\\
 & & 
~\lle^3\,\lhu~; 
~\lhde\,\lhu^4~; 
~\lle\, \lhu^3\;.
\end{eqnarray}

Furthermore, there will be K\"ahler potential terms
\begin{equation}
 K~\supset~X^\dagger\,\left(\frac{k_{\hu\hd}}{M_\mathrm{P}}\hu\,\hd
 +\frac{k_{L\hu\bar\nu}}{M_\mathrm{P}^2}\,L\,\hu\,\bar\nu
 +\frac{k_{QQQL}}{M_\mathrm{P}^3}\,Q\,Q\,Q\,L\right)+\text{h.c.}
\end{equation}
with $X$ denoting the field that breaks supersymmetry, $k_{\hu\hd}$,
$k_{L\hu\bar\nu}$ and $k_{QQQL}$ being coefficients (and the
flavor indices again are suppressed). The $k_{\hu\hd}$ term is nothing but the
famous Giudice--Masiero term \cite{Giudice:1988yz}.

An important feature of this setting is that lepton number is violated at
the quartic level, but bilinear lepton number violating terms are absent. That
is, this model predicts the absence of neutrinoless double $\beta$ decays. On
the other hand, lepton number is not a good symmetry, which might have, for
instance, important implications for the early universe.

Let us also note that the coefficients in the above K\"ahler potential are not
necessarily of order unity. In specific string constructions, these coefficients
can in fact be as large as $\mathcal{O}(10-100)$ due to the presence of copious
heavy states and/or combinatorical factors (cf.\ the discussion in
\cite{Cvetic:1998gv}), enabling realistic predictions for neutrino masses in the
sub--eV range.

\subsubsection{Comments on the $\boldsymbol{\Z4^R}$ symmetries}

Both $\Z{4}^R$ symmetries of \Tabref{tab:Classification} are problematic as 
they allow some $R$ parity violating couplings. In particular, the first
$\Z{4}^R$  allows for bi--linear $R$ parity violation, i.e.\ the $\crep{5}\,\hu$
coupling,  while the second $\Z{4}^R$ admits  the tri--linear $R$ parity 
violating terms 
$\boldsymbol{10}\,\boldsymbol{\overline{5}}\,\boldsymbol{\overline{5}}$.  In
addition, both settings allow for a non--perturbative neutrino bilinear 
$\bar\nu\,\bar\nu$. That is, these symmetries can give us a non--perturbative 
Majorana neutrino mass term, which might be relevant for the construction of 
models realizing a TeV--scale see--saw scenario. Given our previous discussion, 
a straightforward possibility of rectifying this is to amend the settings by 
the residual $\Z4$ symmetry from above (\Tabref{tab:Z4Rextensions}).

\begin{table}[h]
\centerline{\subtable[First $\Z4^R$.]{$\displaystyle
\begin{array}{ccccccccc}
  & q_{\boldsymbol{10}} & q_{\boldsymbol{\overline{5}}} & q_{\hu} & q_{\hd}
 & q_{\theta} & \rho & q_{\bar\nu}\\
 \hline
  \Z4^R & 0 & 0 & 2 & 2 & 1 & 1 & 2 \\
 \Z4 & 1 & 1 & 2 & 2 & 0 & 0 & 1 \\
\end{array}
$}
\qquad
\subtable[Second $\Z4^R$.]{$\displaystyle
\begin{array}{cccccccc}
  & q_{\boldsymbol{10}} & q_{\boldsymbol{\overline{5}}} & q_{\hu} & q_{\hd}
 & q_{\theta} & \rho & q_{\bar\nu}\\
 \hline
 \Z4^R & 2 & 2 & 2 & 2 & 1 & 1 & 0 \\
 \Z4 & 1 & 1 & 2 & 2 & 0 & 0 & 1 \\
\end{array} 
$}
}
\caption{\Z4 extensions of the $\Z4^R$ symmetries of \Tabref{tab:Classification}.}
\label{tab:Z4Rextensions}
\end{table}

The $\Z4^R$ symmetries originally give us two inequivalent Hilbert
superpotential bases, however, amending the settings by the above--mentioned
$\Z4$ symmetry leads to the same basis. Therefore, both $\Z4^R \times \Z4$
symmetries give us the inhomogeneous monomials
\begin{equation}
~\lhde~; 
~\lle\,\bar\nu~;
~\lle\,\lhu^3~;
~\lle^3\,\lhu\;,
\end{equation}
whereas the homogeneous ones are given by
\begin{eqnarray}
 & &
~\bar\nu^4~;
~\hu\,\hd~;
~\lhu\,\bar\nu~;
~\lhu^4~; 
~\lhde\,\lhu\,\lle^3~; \nonumber \\
& &
~\lhde^2~;
~\lle\,\lhde\,\bar\nu~;
~\lle^2\,\bar\nu^2~; \nonumber \\
& &
~\lle^2\,\lhu^2~;
~\lhde\,\lhu^3\,\lle~;
~\lle^4\;.
\end{eqnarray}
As before in our $\Z8^R$ setting, bilinear lepton number violating terms are
absent. In both cases this feature is due to the (anomaly--free non--$R$) \Z4
symmetry, which commutes with \SO{10} for the matter fields and is a consistent
symmetry of the MSSM. Unlike the $R$ symmetries, this symmetry does not forbid
the $\mu$ term nor the dimension five proton decay operators.

\section{Summary}

The MSSM provides a very attractive scheme for physics beyond the standard
model. However, in order to address its shortcomings, one, arguably, has to
impose additional symmetries. Motivated by the structure of matter and the
attractive picture of gauge unification, we have considered symmetries that
commute with \SU{5} in the matter sector. From the requirement of anomaly
freedom it follows that only discrete $R$ symmetries can forbid the $\mu$ term.
We also pointed out that anomaly matching for $R$ symmetries in \SU5 symmetric
models implies the existence of split multiplets below the GUT scale, with the
simplest option being that a pair of Higgs doublets cancels the anomaly mismatch
between the gauginos. Further demanding that a $\mu$ term of the order of the
gravitino mass arises from supersymmetry breaking, i.e.\ either from the
K\"ahler potential or from the non--trivial superpotential VEV in the `hidden
sector', we showed that the Higgs bilinear $\hu\,\hd$ has to carry trivial $R$
charge. We find that discrete $R$ symmetries with these properties automatically
forbid dimension--five proton decay operators once the usual Yukawa couplings
are allowed. Even more, all symmetries appear anomalous such that a holomorphic
$\mu$ term gets induced at the non--perturbative level. That is, demanding
compatibility with the Giudice--Masiero mechanism brings us to the situation in
which a $\mu$ term of the desired magnitude appears even without the
Giudice--Masiero term in the K\"ahler potential.

We then discussed neutrino masses in the emerging MSSM models amended by
discrete $R$ symmetries. Restricting ourselves to flavor--universal Abelian,
i.e.\ $\Z{M}^R$, symmetries we find that, by demanding that the Weinberg operator
$L\,\hu\,L\,\hu$ be allowed, there exists 
only one possible symmetry, namely a
$\Z4^R$ symmetry. Following a different approach, this $\Z4^R$ has also recently
been shown to be the unique anomaly--free symmetry that commutes with
\SO{10}~\cite{Lee:2010gv}. The proof in \cite{Lee:2010gv} assumed that the
charge of the superspace coordinate $\theta$ can always be set 1, which we find
to be too strong a requirement. However, 
we find that, if one is to allow
for arbitrary $\theta$ charges, this only leads to trivial extensions of
$\Z4^R$, such that the uniqueness of $\Z4^R$ still prevails. 

If one requires instead the discrete symmetry to forbid the Weinberg operator,
one can explain small Dirac neutrino masses. In particular,
we successfully obtain a relation between the smallness of Dirac neutrino
Yukawa couplings and the $\mu$ term which is based on 
anomaly--free discrete $R$ symmetries with the above properties. 
Specifically, we find a class of anomaly--free discrete symmetries in which the appealing
relations $\mu\sim\langle\mathscr{W}\rangle/M_\mathrm{P}^2\sim m_{3/2}$ and
$Y_\nu\sim\mu/M_\mathrm{P}$ naturally emerge. 

\subsection*{Acknowledgments}

We would like to thank Maximilian Fallbacher and Hans Peter Nilles for useful
discussions. M.R.\  would like to thank the  UC Irvine, where part of this work
was done, for  hospitality. M.-C.C.\ would like to thank TU M\"unchen, where
part of the work  was done, for hospitality. This work was partially supported
by the DFG cluster  of excellence ``Origin and Structure of the Universe'' and
the Graduiertenkolleg ``Particle Physics at the Energy Frontier of New
Phenomena'' by Deutsche Forschungsgemeinschaft (DFG). P.V.\ is supported by SFB
grant 676.  The work of M.-C.C.\ was supported, in part, by the U.S.\ National
Science  Foundation under Grant No.\ PHY-0970173.   M.-C.C.,  M.R.\ and P.V.\
would like to thank the Aspen Center for Physics for  hospitality and support.
M.-.C.C.\ thanks the Galileo Galilei Institute for Theoretical Physics for the
hospitality.  This research was done in the context of the ERC  Advanced Grant
project ``FLAVOUR''~(267104). 

\appendix

\section{Anomaly coefficients for $\boldsymbol{\Z{M}^R}$ symmetries with
arbitrary $\boldsymbol{q_\theta}$}
\label{sec:AnomalyCoefficients}

The anomaly conditions for discrete $R$ symmetries depend on $q_\theta$.
Consider a $\Z{M}^R$ symmetry, under which the superpotential transforms as
\begin{equation}
 \mathscr{W}~\to~\mathrm{e}^{2\pi\,\I\,q_\mathscr{W}/M}\,\mathscr{W}
\end{equation}
with $q_\mathscr{W}=2q_\theta$ (such that $\int\!\D^2\theta\,\mathscr{W}$ is
invariant). 
Superfields
$\Phi^{(f)}=\phi^{(f)}+\sqrt{2}\,\theta\psi^{(f)}+\theta\theta\,F^{(f)}$
transform as
\begin{equation}
 \Phi^{(f)}~\to~\mathrm{e}^{2\pi\,\I\,q^{(f)}/M}\,\Phi^{(f)}\;.
\end{equation}
Correspondingly, the fermions transform as
\begin{equation}
 \psi^{(f)}~=~\mathrm{e}^{2\pi\,\I\,(q^{(f)}-q_\theta)/M}\,\psi^{(f)}\;.
\end{equation}
The anomaly coefficients hence read (cf.\ \cite[Appendix~B]{Lee:2011dy}, where
the anomaly coefficients for the special case $q_\theta=1$ are shown)
\begin{subequations}
\begin{eqnarray}
\label{eq:AGGZN}
 A_{G-G-\Z{M}^R}
 & = &
 \sum_f \ell(\boldsymbol{r}^{(f)})\cdot(q^{(f)}-q_\theta)
 +q_\theta\,\ell(\adj G)
 \;,\label{eq:A_G-G-ZNR}\\
 A_{\U1-\U1-\Z{M}^R}
 & = &
 \sum_f (Q^{(f)})^2\,\dim(\boldsymbol{r}^{(f)})\cdot(q^{(f)}-q_\theta)
 \;,\label{eq:A_U1-U1-ZNR}\\
 A_{\mathrm{grav}-\mathrm{grav}-\Z{M}^R}
 & = &
 -21\,q_\theta+q_\theta\,\sum_G \dim(\adj G)
 +\sum_f \dim(\boldsymbol{r}^{(f)})\cdot(q^{(f)}-q_\theta)
 \;.     
 \label{eq:A_grav-grav-ZNR}
\end{eqnarray}
\end{subequations}
Here $q^{(f)}$ denote the $\Z{M}^R$ charges of the superfields, the
charges of the corresponding fermions are shifted by $q_\theta$,
$q_{\psi^{(f)}}=q^{(f)}-q_\theta$. In \Eqref{eq:A_G-G-ZNR}, 
$\ell(\boldsymbol{r}^{(f)})$ denotes the Dynkin index of representation 
$\boldsymbol{r}^{(f)}$ normalized to $\ell(\boldsymbol{N}) = \frac{1}{2}$ 
for the fundamental representation $\boldsymbol{N}$ of $\SU{N}$ and 
$\ell(\adj G)=c_2(G)$ represents the contribution from the gauginos, 
i.e.\ $\ell(\adj \SU{N})= N$. The first
and second terms on the right--hand side of \Eqref{eq:A_grav-grav-ZNR}
represent the contributions from the gravitino and gauginos. 

\section{Green--Schwarz anomaly cancellation and anomaly universality}
\label{app:GS}

In this Appendix, we discuss the discrete Green--Schwarz (GS) anomaly
cancellation mechanism, following \cite{Lee:2011dy}. We start by reviewing the GS mechanism for a continuous
\U1 symmetry in \ref{app:GSU(1)}. In \ref{app:DiscreteGS} we discuss the
discrete version while \ref{app:AnomalyUniversality} is dedicated to the
discussion of anomaly universality.

\subsection{Anomaly cancellation for `anomalous U(1)' symmetries}
\label{app:GSU(1)}

We start by discussing the mixed anomaly coefficients $G-G-\U1_\mathrm{anom}$ for a simple
gauge group $G$. There will be an axion $a$ which couples to the field strength
of $G$ via
\begin{equation}\label{eq:Laxion1}
 \mathscr{L}_{\mathrm{axion}}
 ~\supset~
 \frac{a}{8} F^{b} \widetilde{F}^b 
\;.
\end{equation}
A possible prefactor can be absorbed in the normalization of $a$, which we do
not specify here.
Consider now the gauge transformation
\begin{equation}\label{eq:InfinitesimalTrafo1}
 \psi^{(f)}~\to~\mathrm{e}^{\I\,\alpha(x)\,Q_\mathrm{anom}^{(f)}}\,\psi{(f)}\;,
\end{equation}
where $\psi^{(f)}$ ($1\le f\le F$) denotes the fermions of the theory and
$Q_\mathrm{anom}^{(f)}$ their charges. The crucial property of the axion $a$ is
that it shifts under \eqref{eq:InfinitesimalTrafo1} as 
\begin{equation}\label{eq:AxionShift}
 a~\to~a+\frac{1}{2}\delta_\mathrm{GS}\,\alpha(x)\;.
\end{equation}

We can now fix the the Green--Schwarz coefficient $\delta_\mathrm{GS}$ from the
requirement of invariance of the full quantum theory. It follows from
\eqref{eq:AxionShift} that, under a $\U1_\mathrm{anom}$ transformation with
parameter $\alpha$, the axionic Lagrangean shifts by
\begin{equation}
  \Delta\mathscr{L}_{\mathrm{axion}}  ~=~
  -\frac{\alpha}{16}\delta_\mathrm{GS}\, F^{b} \widetilde{F}^b
  \;.
\end{equation}
The Green--Schwarz term $\delta_\mathrm{GS}$ can now be inferred by
demanding that the transformation of the axion $a$ cancels the
anomalous variation of the path integral measure
\cite{Fujikawa:1979ay,Fujikawa:1980eg}. The latter can be
absorbed in a change of the Lagrangean
\begin{eqnarray}
 \Delta\mathscr{L}_{\text{anomaly}} &= &  
 \frac{\alpha}{32\pi^2} F^{b} \widetilde{F}^b  \, A_{G-G-\U1_\mathrm{anom}}
\;.
\end{eqnarray}
The coefficient $A$ is the anomaly coefficient, given by
\begin{equation}\label{eq:anomaly_coefficients}
 A_{G-G-\U1_\mathrm{anom}} ~ = ~\sum_{\rep{r}^{(f)}}
 \ell(\rep{r}^{(f)}) \, Q_\mathrm{anom}^{(f)} \;,
\end{equation}
where the sum runs over all irreducible (fermionic) representations
$\rep{r}^{(f)}$ of $G$, $\ell(\boldsymbol{r}^{(f)})$ denotes the Dynkin
index of $\boldsymbol{r}^{(f)}$ and $Q_\mathrm{anom}^{(f)}$ is the
$\U1_{\mathrm{anom}}$ charge.

The axion shift allows us to cancel the $G-G-\U1_{\mathrm{anom}}$ anomaly by
demanding $\Delta\mathscr{L}_{\mathrm{anomaly}} +
\Delta\mathscr{L}_{\mathrm{axion}}=0$. This fixes the Green--Schwarz constant to
be
\begin{equation}\label{eq:AnomalyUniversalityU(1)}
 2\pi^2\,\delta_{\mathrm{GS}}
 ~=~
 A_{G-G-\U1_\mathrm{anom}} \;.
\end{equation}

\subsection{Discrete Green--Schwarz mechanism}
\label{app:DiscreteGS}

The Green--Schwarz mechanism also works if we replace $\U1_{\mathrm{anom}}$ by a
discrete \Z{M}. In this case the parameter $\alpha$ is no longer continuous but
$\alpha=\frac{2\pi n}{M}$ with some integer $n$. Of course, there is no gauge
field associated with the \Z{M}. The discussion then goes as in the previous
subsection. The discrete Green--Schwarz constant is now defined in such a way
that under the $\Z{M}$ transformation of fermions 
\begin{equation}
 \psi^{(f)}~\to~\mathrm{e}^{-\I\,\frac{2\pi}{M}\,q^{(f)}}\,\psi^{(f)}
\end{equation}
the axion shifts according to
\begin{equation}\label{eq:DiscreteShiftDilaton}
 a~\to~a+\frac{1}{2}\Delta_\mathrm{GS}\;,
\end{equation}
where $\Delta_\mathrm{GS}$ is fixed only modulo $\eta$,
\begin{equation}\label{eq:DeltaGS}
 \pi\,M\,\Delta_\mathrm{GS}
 ~ \equiv ~
 A_{G-G-\Z{M}}
 \mod \eta
 \;.
\end{equation}
The anomaly coefficients can be obtained from \Eqref{eq:anomaly_coefficients} by
replacing the $\U1_{\mathrm{anom}}$ charges $Q_\mathrm{anom}^{(m)}$ by the \Z{M}
charges $q^{(m)}$. 

\subsection{Multiple gauge groups and ``anomaly universality''}
\label{app:AnomalyUniversality}

Let us now discuss the case of multiple gauge groups $G_i$. In heterotic string
models very often the $\U1_\mathrm{anom}$, \Z{N} or $\Z{M}^{R}$ anomaly
coefficients fulfill certain universality relations, 
\begin{equation}\label{eq:AnomalyUniversalityGi}
 A_{G_i-G_i-H}~=~\rho
\end{equation}
for all $i$ and in this section $H$ denotes either $\U1_\mathrm{anom}$,
\Z{N} or $\Z{M}^{R}$. We will refer to \eqref{eq:AnomalyUniversalityGi} as
``anomaly universality''. In a recent paper \cite{Ludeling:2012cu} it has been
pointed out correctly that this may not necessarily be the case in general. That
is, the anomaly  universality \eqref{eq:AnomalyUniversalityGi} is not a direct
consequence of GS  anomaly cancellation.

In detail, multiple gauge groups $G_i$ in general do allow us to introduce 
different couplings $c_i$ of the axion $a$ to the various field strengths,
\begin{equation}\label{eq:Laxion2}
 \mathscr{L}_{\mathrm{axion}}
 ~\supset~
 \sum\limits_i  c_i\,\frac{a}{8} F_i^{b} \widetilde{F}_i^b \;.
\end{equation}
The requirement that under an $H$ transformation the contribution from 
the path integral measure gets cancelled by the discrete shift of the axion 
then implies that
\begin{equation}\label{eq:AnomalyUniversality2}
 2\pi^2\,c_i\,\delta_{\mathrm{GS}}
 ~=~
 A_{G_i-G_i-H} 
\end{equation}
for all $i$. That means that the $c_i$ coefficients can be chosen in such a way
that the transformation of the path integral measure gets cancelled for each
$G_i$ gauge factor separately. In particular, one finds (in agreement with
\cite{Ludeling:2012cu})  that in general the mixed $A_{G_i-G_i-H}$ do not need
to be universal.

However, if this was the case in a given model, one would spoil the beautiful
picture of MSSM  gauge coupling unification. Let us spell out the argument in
some more detail.  In supersymmetry, the Lagrangean \eqref{eq:Laxion2} implies
that there are  couplings between the superfield $S$ which contains the axion, 
$S|_{\theta=0}=s+\I\,a$, and the supersymmetric field strengths $W^{(i)}$
associated to the gauge group factors $G_i$, i.e.
\begin{equation} 
 \mathscr{L}_{\mathrm{axion}}
 ~\supset~
 \sum_i\int\!\D^2\theta\, \frac{c_i}{8}\,S\,W_\alpha^{(i)}W^{(i)\,\alpha}\;.
\end{equation}
Once the real part of $S$ acquires a VEV this will give rise to a non--universal
change of the gauge couplings unless the $c_i$ coefficients are all equal for
the SM gauge group factors $G_i = $ $\SU3_\mathrm{C}$, $\SU2_\mathrm{L}$  and
$\U1_Y$. That is, anomaly universality is also required in order not to spoil 
the beautiful picture of MSSM gauge coupling unification.

Furthermore, there might be model--dependent reasons why the $A_{G_i-G_i-H}$ can
be universal, for instance if all $G_i$ come from a (for instance grand unified)
simple gauge group, as we assume in the main body of this paper. Then the term
\begin{equation}
\label{eq:AnomalyUniversalityFromGUT}
 \mathscr{L}_{\mathrm{axion}}
 ~\supset~
 a\,F_\mathrm{GUT}^{b} \widetilde{F}_\mathrm{GUT}^b 
\end{equation}
is obviously gauge invariant. Hence, the anomalies need to be universal at 
least at the GUT level, as discussed around \Eqref{eq:AnomalyMatching2}.

How could this universality possibly be broken? One may now worry about 
additional terms of the form
\begin{equation}
 \mathscr{L}_{\mathrm{axion}}
 ~\supset~
 a\,
 \left(
 	\frac{\Phi_\mathrm{GUT}}{M}\,F_\mathrm{GUT}\widetilde{F}_\mathrm{GUT}
 \right)\;,
\end{equation}
where the operator $\Phi_\mathrm{GUT}$ furnishes a non--trivial GUT
representation (such as a $\rep{24}$--plet of \SU5) and the parentheses denote a
non--trivial contraction of the group indices.\footnote{The relative 
coefficients $c_i$ of the axion coupling to the three $F_i\,\widetilde{F}_i$ 
terms of the standard model originating from $\Phi_\mathrm{GUT}$ can be inferred
from \cite{Huitu:1999vx}.} However, at the GUT level, i.e.\ for a trivial
$\Phi_\mathrm{GUT}$ VEV, such a term can not cancel the transformation  of the
path integral measure by a shift transformation of the axion $a$. In other
words, it is not allowed by the symmetries of the action if we require that $a$
shifts. Hence, these operators can not break anomaly universality.

However, there is a second possibility. In higher--dimensional, e.g.\ in 
orbifold GUT type, models there can be localized terms which do not respect the
GUT symmetry. That is, in settings where the GUT symmetry is broken locally in
some regions of compact space such as orbifold fixed points, anomaly
non--universality can arise. After integrating over compact space in order to
derive the four--dimensional effective action one can indeed arrive at
non--universal couplings $c_i$ of the axion to the three $F\,\widetilde{F}$
terms of the standard model. Still, as discussed before, if one is not to spoil
the beautiful picture of MSSM gauge coupling unification, the $A_{G_i-G_i-H}$
coefficients need to  be universal and these localized  contributions have to be
avoided. One  possibility to avoid them is ``non--local  GUT breaking'' in extra
dimensions,  which has been argued to yield the most  appealing scenarios of
precision gauge
unification~\cite{Hebecker:2004ce,Trapletti:2006xv,Anandakrishnan:2012ii}. In 
such scenarios, the localized dangerous GUT--breaking operators do not exist and
hence the anomalies are universal.\footnote{In Abelian orbifold models such
operators can only stem from localized fluxes, which are Abelian (i.e.\ \U1)
fluxes. Hence, the $A_{G_i-G_i-H}$ coefficients coincide for all non--Abelian
factors  $G_i$ from each $\E{8}$ in such models.  This is also in agreement
with  \cite{Ludeling:2012cu} where it is found that, in compactifications of the
heterotic  $\E{8}\times\E{8}$ string on blown--up orbifolds with Abelian
fluxes,  non--Abelian anomalies  of each $\E{8}$ factor are still universal.
Since the  relevant assumption in \Secref{sec:AnomalyConstraints} needed to
prove the  uniqueness of $\Z4^R$ (see also \Secref{sec:A1notneeded}) is that
$A_{\SU2_\mathrm{L}-\SU2_\mathrm{L}-\Z{M}^R}$ and 
$A_{\SU3_\mathrm{C}-\SU3_\mathrm{C}-\Z{M}^R}$ coincide, the uniqueness of
$\Z4^R$ is also given in such constructions.}

Let us also comment on another statement in \cite{Ludeling:2012cu}. First, we
would like to point out that the number of axions is not related to anomaly 
universality. Specifically, in the presence of multiple axions, which are 
available in heterotic compactifications 
\cite{Blumenhagen:2005ga,GrootNibbelink:2007ew},  one would
have to define how they transform under a $\U1_\mathrm{anom}$ (or discrete)
transformation. Since there is only one such transformation, this allows us to
identify one unique linear combination of axions, called $a$ as in our
discussion above, which shifts while the other `would--be axions' stay inert.
Therefore, the number of axions is not related to the question of anomaly
(non--)universality.

Furthermore, the authors of \cite{Ludeling:2012cu}  argue that the anomalies
cannot be universal both before and after doublet--triplet splitting. We
disagree with this statement. First of all, `before doublet--triplet splitting',
i.e.\ before GUT breaking, there  are more states around which contribute to the
anomalies and anomaly  universality follows from gauge invariance under the GUT
group, see  \Eqref{eq:AnomalyUniversalityFromGUT}. Moreover, in the absence of 
localized GUT breaking terms, if the anomaly coefficients are universal at the 
GUT level, where the contributions of extra states have to be taken into 
account, they should also be so in the MSSM. This is, again, nothing but 't
Hooft anomaly matching (see \Secref{sec:AnomalyMatching}). In fact, at the GUT
level there is just one (unified) gauge group, such that universality is
trivial.

\section{The Hilbert superpotential bases for models with
$\boldsymbol{\Z{12}^{R}}$ symmetries}
\label{app:Hilbert}

In Section \ref{sec:Dirac}, we discuss several $\Z{M}^{R}$ symmetries that forbid
neutrino masses perturbatively and also present the Hilbert superpotential basis
for a model with a $\Z{8}^{R}$ symmetry and two $\Z4^R$ symmetries amended by
an extra \Z4 factor. In this appendix we provide further examples based on the
$\Z{12}^{R}$ symmetries. As we have already stated above, every possible
superpotential term $\mathscr{M}$ contains only one inhomogeneous monomial and
an arbitrary combination of homogeneous monomials \cite{Kappl:2011vi}, i.e.
\begin{equation}
 \mathscr{M} ~=~ \mathscr{M}_{\mathrm{in}}^{(i)}\, \prod_{j=1}\, \left(\mathscr{M}_{\mathrm{hom}}^{(j)}\right)^{\eta_{j}}\, \qquad \mathrm{with}\, \qquad \eta_{j}\, \in\, \mathbbm{N}\;,     
\end{equation}
where $\mathscr{M}_{\mathrm{in}}^{(i)}$ is an inhomogeneous and $\mathscr{M}_{\mathrm{hom}}^{(j)}$ a homogeneous monomial. \newline

In Section \ref{sec:Dirac} we list three examples which have a $\Z{12}^{R}$ symmetry. 
As we will see in the following, the three sets of monomials differ.  Hence, the
three $\Z{12}^{R}$ symmetries are inequivalent. The first symmetry has the
charge assignment
\begin{equation}
 \left(\begin{array}{ccccccc}
q_{\boldsymbol{10}} & q_{\boldsymbol{\overline{5}}} & q_{\hu} & q_{\hd}
 & q_{\theta} & \rho & q_{\bar\nu}
\end{array}\right) 
~=~ \left(\begin{array}{ccccccc}
           1 & 9 & 4 & 8 & 3 & 3 & 11
          \end{array}\right)\;,
\end{equation}
which leads to the inhomogeneous monomials
\begin{eqnarray}
 & &
~\lhde~;
~\lhu^6~;
~\bar\nu^6~;
~\lle\, \bar\nu~; \nonumber \\
 & &
~\lle^6~;
~\lle^4\,\lhu^2~;
~\lle^2\, \lhu^4\;,
\end{eqnarray}
whereas the homogeneous ones are given by
\begin{eqnarray}
 & &
~\lle^{12}~;
~\lhu^{12}~;
~\hu\,\hd~;
~\left(L\,\hu\right)\, \bar\nu~; 
~\lle\, \lhde\, \bar\nu~; \nonumber \\
 & &
~\bar\nu^{12}~;
~\lhde\,\lhu^6~;
~\lhde^2~;
~\lle^7\, \bar\nu~; \nonumber \\
 & &
~\lle\, \lhu^5~;
~\lle^6\, \lhde~;
~\lhde\, \bar\nu^6~; \nonumber \\
 & &
~\lle^2\, \bar\nu^2~;
~\lhde\, \lle^4\, \lhu^2~;
~\lle^5\, \lhu ~; \nonumber \\
& &
~\lhde\, \lle^2\, \lhu^4~; 
~\lle\, \bar\nu^7 ~; 
~\lle^3\, \lhu^3\;.
\end{eqnarray}
The second $\Z{12}^{R}$ symmetry has the charges
\begin{equation}
 \left(\begin{array}{ccccccc}
q_{\boldsymbol{10}} & q_{\boldsymbol{\overline{5}}} & q_{\hu} & q_{\hd}
 & q_{\theta} & \rho & q_{\bar\nu}
\end{array}\right) 
~=~ \left(\begin{array}{ccccccc}
           2 & 6 & 2 & 10 & 3 & 3 & 4
          \end{array}\right)\;,
\end{equation}
which gives us for the inhomogeneous monomials
\begin{eqnarray}
 & &
~\lhde~;
~\lle^3~;
~\lle \bar\nu~;
~\lle\, \lhu^2\;,
\end{eqnarray}
and for the homogeneous monomials
\begin{eqnarray}
 & &
~\lle^6~;
~\lle^4\, \bar\nu~;
~\lhde\, \lle^3~;
~\hu\,\hd~; \nonumber \\
& &
~\left(L\,\hu\right)\, \bar\nu~;
~\bar\nu^3~;
~\lhu^3~;
~\lhde\, \lhu^2\, \lle~; \nonumber \\
& &
~\lle^2\, \lhu~;
~\lhde^2~;
~\lle^2\, \bar\nu^2~;
~\lhde\, \lle\, \bar\nu\;. 
\end{eqnarray}
The last $\Z{12}^{R}$ symmetry has 
\begin{equation}
 \left(\begin{array}{ccccccc}
q_{\boldsymbol{10}} & q_{\boldsymbol{\overline{5}}} & q_{\hu} & q_{\hd}
 & q_{\theta} & \rho & q_{\bar\nu}
\end{array}\right) 
~=~ \left(\begin{array}{ccccccc}
           4 & 0 & 10 & 2 & 3 & 3 & 2
          \end{array}\right)\;,
\end{equation}
as its charge assignment, with these we get the inhomogeneous monomials
\begin{eqnarray}
 & &
~\bar\nu^3~;
~\lhde~;
~\lle\, \bar\nu~;
~\lhu^3~;
~\lle^2\,\lhu\;,
\end{eqnarray}
and the following homogeneous ones
\begin{eqnarray}
 & &
~\lhu^6~;
~\lhde\,\lhu^3~;
~\hu\,\hd~;
~\lhde\,\lle^2\,\lhu~; \nonumber \\
& &
~\lhu\,\bar\nu~;
~\lle^3~;
~\lle\,\lhu^2~;
~\lhde^2~; \nonumber \\
& &
~\lhde\, \bar\nu^3~;
~\lhde\,\lle\,\bar\nu~;
~\lle\,\bar\nu^4~;
~\bar\nu^6~; \nonumber \\
& &
~\lle^2\, \bar\nu^2\;.
\end{eqnarray}

\bibliography{Orbifold}
\addcontentsline{toc}{section}{Bibliography}
\bibliographystyle{NewArXiv} 
\end{document}